\documentstyle [12pt,a4p,epsfig,amsmath,multicol]{article}
\begin{document}

\begin{center} 
{\normalsize\bf The Meaning of Coherence in Weak Decay Processes:
 `Neutrino Oscillations' Reconsidered}
\end{center}
\vspace*{0.6cm}
\centerline{\footnotesize J.H.Field}
\baselineskip=13pt
\centerline{\footnotesize\it D\'{e}partement de Physique Nucl\'{e}aire et 
 Corpusculaire, Universit\'{e} de Gen\`{e}ve}
\baselineskip=12pt
\centerline{\footnotesize\it 24, quai Ernest-Ansermet CH-1211Gen\`{e}ve 4. }
\centerline{\footnotesize E-mail: john.field@cern.ch}
\baselineskip=13pt
 
\vspace*{0.9cm}
\abstract{ A Feynman path integral analysis of a two-neutrino-flavour electron
 appearence experiment following pion decay at rest recovers the standard oscillation
 phase, revealing an important contribution from the decay amplitude of the pion
 as well as an error in previous similar calculations by the present author.
 In the calculation, path amplitudes for different neutrino mass eigenstates add
 coherently, but no putative `neutrino flavour eigenstates' are invoked.
 It is shown that the coherent production of the latter states is incompatible with
 the measured values of $\Gamma(\pi \rightarrow {\rm e}\nu)/\Gamma(\pi \rightarrow \mu \nu)$
 and the PMNS matrix elements. Application of the path integral approach to other
 two-path quantum interference experiments is compared with that to neutrino
 oscillations, and other treatments of the latter in the literature are discussed.}
\vspace*{0.9cm}
\normalsize\baselineskip=15pt
\setcounter{footnote}{0}
\renewcommand{\thefootnote}{\alph{footnote}}
\newline
{\it Keywords ;} Quantum Mechanics,
Neutrino Oscillations.
\newline

\vspace*{0.4cm}


  In the Standard Electroweak Model (SEM), the coupling of a charged lepton, $\ell_i$, of
  generation $i$ and a neutrino mass eigenstate $\nu_j$, of generation $j$  to the
   W-boson is proportional to $ij$th component of the leptonic
 charged current:
 \begin{equation}
 J_{\mu}(CC)^{lept} = \sum_{i,j} \overline{\psi}_{\ell_i} \gamma_{\mu}(1-\gamma_5)U_{ij}\psi_{\nu_j}
 \end{equation}
  where $U_{ij}$ is the Pontecorvo-Maki-Nakagawa-Sakata (PMNS)~\cite{PC,MNS}
  charged lepton flavour/ neutrino mass
  mixing matrix. Table 1 shows an early estimate~\cite{GGN} of the elements of this matrix 
 from experimental measurements of atmospheric and solar neutrino oscillations.
 The non-diagonal nature of this matrix gives evidence for strong violation
 of generation number (or lepton flavour) by $ J_{\mu}(CC)^{lept}$. Conservation of
 generation number corresponds to a diagonal PMNS matrix with $\nu_1 = \nu_e$,
 $\nu_2 = \nu_{\mu}$ and  $\nu_3 = \nu_{\tau}$. This is the conventional massless
  neutrino scenario. With massive neutrinos and a non-diagonal PMNS matrix the leptonic charged current
  (1) contains only the mass eigenstates $\nu_j$ of mass $m_j$ 
      so that in this case the `flavour eigenstates': `$\nu_e$',
 `$\nu_{\mu}$' and `$\nu_{\tau}$' do not exist. That is, they do
  not appear in the amplitude for any physical process of the SEM. 

 \begin{table}
   \begin{center}
   \begin{tabular}{|c|c|c|c|} \hline  
       $j$  & 1 ($\nu_1$)  & 2 ($\nu_2$) &  3 ($\nu_3$) \\
   \hline \hline
      $i$  &   &   &    \\ \cline{1-1}
     1 ($e$)  & $0.79 \pm 0.12$   &  $0.57 \pm 0.16$ & $0.1 \pm 0.1$  \\
     2  ($\mu$) & $-0.45 \pm 0.25$  & $0.49 \pm 0.28$ & $0.69 \pm 0.18$  \\
    3  ($\tau$) & $0.34 \pm 0.29$    & $-0.60 \pm 0.23$   &  $0.67 \pm 0.18$ \\
  \hline
  \end{tabular}
   \caption[]{ Values of the MNS lepton flavour/neutrino mass mixing
   matrix $U_{ij}$ as derived from solar and atmospheric neutrino
   oscillation data~\cite{GGN}. }
  \end{center}
  \end{table} 

   \par An analysis of two-flavour neutrino oscillations, following pion decay at rest, within Feynman's
      path integral formulation of quantum  mechanics, is now presented. In this case, without
     loss of generality, the PMNS matrix elements are assumed to be real numbers. A close comparison with another 
      two-path experiment, the Young double slit, in physical optics, will be made. This comparison will
     reveal an incorrect physical postulate in previous treatments~\cite{JHFEPJC1,JHFEPJC2,JHFAP} of neutrino
    oscillations by the present author. Following the sequential factorisation law~\cite{JHFAP} for
     constructing path amplitudes, each such amplitude in a two-flavour neutrino oscillation experiment or a two path
     experiment in photonic physical optics, will be the product of the following amplitudes:
\begin{itemize}
\item[(i)] The amplitude to produce the source particle.
\item[(ii)] The decay amplitude of the source particle into a final state containing a neutrino mass eigenstate or a photon
\item[(iii)] The space-time propagator of the neutrino or the photon.
\item[(iv)] The amplitude of the process by which the  neutrino or photon is detected.
 \end{itemize}
  The superposition principle for path amplitudes~\cite{FeynRMP,JHFAP} requires that if, and only if, the
  path amplitudes have the same initial and final states they must be added coherently, i.e. the amplitudes,
   not the modulus squared of the amplitudes, must be summed. This coherence condition is
   completely different to the hypothesis to be discussed below, that is the basis of `standard'
   neutrino oscillation phenomenology that a `neutrino flavour eigenstate' that is a superposition
   of neutrino mass eigenstates, is produced at some fixed time.
   The production amplitude in (i) is common to both path amplitudes and therefore
   contributes only an overall multiplicative factor to the oscillation probability or interference pattern.
   For the neutrino oscillation experiment, the initial state of the path amplitudes is that of the pion at the instant of its creation.  
   The amplitude in (ii) is a function of the time interval, $t_{{\rm P}}$, after the source particle is
   created, at which the decay occurs~\cite{DiracQM,JHFAP}:
   \begin{equation}
  \langle f|i\rangle_{t_{{\rm P}}}  = \exp \left[-i\frac{(E_i-E_f)t_{{\rm P}}}{\hbar}\right] \langle f|i\rangle_0
    \end{equation}
    where $\langle f|i\rangle_0$ and $\langle f|i\rangle_{t_{{\rm P}}}$ are the transition amplitudes at time
    zero and $t_{{\rm P}}$ respectively. The suffix 'P' stands for 'Production' (of the neutrino or photon).
    In the formula (2) it is assumed that the lifetime of the source particle is much greater than
    the difference between the times-of-flight of the neutrinos or photons in the two paths.
    The source particle is produced at time zero in both path amplitudes. In the physical optics application
    of (2) $E_i$ and  $E_f$ are the energies of atomic states and $E_i-E_f= E_{\gamma}$ where $E_{\gamma}$
    is the photon energy. The conceptual error in~\cite{JHFEPJC1,JHFEPJC2,JHFAP} and 
     earlier versions of the present paper~\cite{JHFMCNO} was to replace the amplitude (ii) by the space-time
     propagator of the source particle. Since the same laws of physics should apply for both neutrino
     oscillations and physical optics this corresponds, in the latter case, to replacing $E_{\gamma}$
     by $M_i c^2$ where $M_i$ is the mass of the unstable source atom! The `photon wavelength' 
     governing interference effects would then be smaller by the factor $E_{\gamma}/(M_i c^2)$
     as compared with the value in the classical wave theory of light~\cite{JHFAP} --- evidently at variance
      with experiment.
     \par Since the space-time propagator of a free particle has the phase: $(rp-Et)/\hbar$~\cite{JHFAP}, and 
    for a photon $c = r/t = E/p$, the propagator phase vanishes~\cite{FeynQED} so that the path amplitude 
    phase resides entirely in (ii) and is given by (2) with $E_i-E_f= E_{\gamma}$. For the case of neutrino
    oscillations (2) holds with $E_i-E_f= E_{\nu_j} \equiv E_j$, whereas the phase of the neutrino propagator
     is~\cite{JHFEPJC1,JHFAP}: $-m_{j}c^2 \tau_{{\rm F}}/\hbar$ where $\tau_{{\rm F}}$ is the time-of-flight
     of the neutrino in its rest frame.
       \par The final state of both path amplitudes is that of the detection process described
      by the amplitude (iv).
     \par Consider now production of the neutrino mass eigenstates $\nu_1$ or $\nu_2$ in the two-body
       decay at rest of a positively charged pion: $\pi^+ \rightarrow \mu^+ \nu_1$ or $\mu^+ \nu_2$. A 
       `neutrino oscillation' effect is manifested by detection of a neutrino via the
         processes: $(\nu_1,\nu_2) n \rightarrow e^- p$ at a fixed distance, $L$, from the source. In
       units with $\hbar = c =1$ the path amplitude for the mass eigenstate $\nu_j$ is, up to
       a overall multiplicative constant~\cite{JHFEPJC1,JHFAP}:
     \begin{equation}
      A^j_{{\rm e}\mu \pi}(t^j_{{\rm P}}) = U_{{\rm e}j}\langle {\rm e}^-|\nu \rangle \exp\left[-i \frac{m_j^2L}{p_j}\right]
       \exp\left\{-iE_j t^j_{{\rm P}}\right\} U_{\mu j}\langle \nu \mu^+|\pi^+\rangle 
      \end{equation}
         where the `reduced' decay and scattering amplitudes $\langle \nu \mu^+|\pi^+\rangle$ and 
         $\langle{\rm e}^-|\nu\rangle$ are defined according to the relations:
       \begin{equation}    
        \langle \nu_j \mu^+|\pi^+\rangle_0 \equiv U_{\mu j} \langle \nu \mu^+|\pi^+\rangle,~~~~
        \langle{\rm e}^-|\nu_j\rangle \equiv U_{{\rm e} j}\langle{\rm e}^-|\nu \rangle.
  \end{equation}
    The amplitudes (ii)-(iv) are written as factors from right to left on the right
     side of (3). Extracting the the modulus and phase of the path amplitude in (3):
       \begin{equation}  
      A^j_{{\rm e}\mu\pi}(t^j_{{\rm P}}) = U_{{\rm e}j} U_{\mu j}
      |\langle {\rm e}^-|\nu \rangle|| \langle \nu \mu^+|\pi^+\rangle| e^{i(\phi_j+\phi_0)}
         \equiv A^{0 j}_{{\rm e} \mu \pi} e^{i(\phi_j+\phi_0)}
        \end{equation}
       where $\phi_0$ is a possible flavour-independent phase associated with the amplitudes
     $\langle \nu \mu^+|\pi^+\rangle$ and $\langle{\rm e}^-|\nu \rangle$ and
    \begin{equation}
    \phi_j = -\left(\frac{m_j^2 L}{p_j}+E_j t^j_{{\rm P}}\right). 
       \end{equation}
      Quantum mechanical superposition of the path amplitudes~\cite{FeynRMP,JHFAP} gives, for the
      probability to detect an electron:
    \begin{equation}
       P_{{\rm e}\mu \pi} = |A^1_{{\rm e}\mu \pi}+A^2_{{\rm e}\mu\pi}|^2
       = ( A^{0 1}_{{\rm e} \mu \pi})^2 +( A^{0 2}_{{\rm e} \mu \pi})^2+ 2 A^{0 1}_{{\rm e} \mu \pi}
           A^{0 2}_{{\rm e} \mu \pi} \cos(\phi_1-\phi_2).
         \end{equation}
        Introducing the neutrino production time difference: $2 \Delta t_{{\rm P}}$ and the mean neutrino
      production time $\bar{t}_{{\rm P}}$:
         \begin{equation}
           \Delta t_{{\rm P}} \equiv \frac{t^1_{{\rm P}}-t^2_{{\rm P}}}{2},~~~~
            \bar{t}_{{\rm P}} \equiv \frac{t^1_{{\rm P}}+t^2_{{\rm P}}}{2}
       \end{equation}
     enables the phase difference between the two path amplitudes to be written, using (6), as:
      \begin{equation}
     \Delta \phi_{12} \equiv \phi_1-\phi_2 = \left(\frac{m_2^2}{p_2}-\frac{m_1^2}{p_1}\right)L
           -(E_1+E_2)\Delta t_{{\rm P}}+(E_2-E_1) \bar{t}_{{\rm P}}. 
      \end{equation}
       If $t_f^1$, $t_f^2$ are the times-of-flight of $\nu_1$ and  $\nu_2$ between production
    and detection at the common time $t_{{\rm D}}$ then
        \begin{equation}
         t_{{\rm D}} = t^1_{{\rm P}}+t_f^1 = t^2_{{\rm P}}+t_f^2
        \end{equation}    
       so that    
           \begin{equation}
          t^1_{{\rm P}}-t^2_{{\rm P}} = 2 \Delta t_{{\rm P}} =t_f^2-t_f^1 
        \end{equation}  
        and since
      \begin{equation}
       t_f^j =\frac{L}{v_j} = \frac{E_j L}{p_j}~~~j =1,2  
  \end{equation}
      it follows that
         \begin{equation}          
 \Delta t_{{\rm P}} = \frac{L}{2}\left(\frac{E_2}{p_2}-\frac{E_1}{p_1}\right). 
 \end{equation}
   Exact relativistic two-body kinematics of the decay process $\pi \rightarrow \mu \nu_j$ gives:
   \begin{equation}  
     E_j = \frac{m_{\pi}^2-m_{\mu}^2}{2 m_{\pi}}+ \frac{m_j^2}{2 m_{\pi}}
      \equiv E_{\nu} + \frac{m_j^2}{2 m_{\pi}}~~~~j= 1,2,   
 \end{equation}
    \begin{equation}
        E_2-E_1 = \frac{\Delta m_{21}^2}{2 m_{\pi}},~~~~\Delta m_{21}^2 \equiv m_2^2- m_1^2. 
    \end{equation}
     Since
    \begin{equation}
      p_j = E_j- \frac{m_j^2}{2 E_{\nu}}+{\rm O}(m_j^4)
   \end{equation}
     (13) gives
    \begin{equation}
  \Delta t_{{\rm P}} = \frac{\Delta m_{21}^2 L}{4 E_{\nu}^2}+{\rm O}(m_j^4)
    \end{equation}
    Combining (13)-(17) and (9) gives:
      \begin{equation}
    \Delta \phi_{12} = \frac{\Delta m_{21}^2 }{2 E_{\nu}}\left[L+\frac{E_{\nu} c\bar{t}_{{\rm P}}}{E_{\pi}}\right]
     +{\rm O}(m_j^4)
      \end{equation}
     where both terms in the square bracket are of dimension [L]. Inserting the values of the PMNS matrix elements
     in terms of the two-flavour mixing angle $\theta_{12}$:
    \begin{equation}
 \left(\begin{array}{cc}
    U_{e 1} & U_{e 2} \\
  U_{\mu 1} & U_{\mu 2}
 \end{array} \right) = 
  \left(\begin{array}{cc}
    \cos \theta_{12} &  \sin \theta_{12} \\
   - \sin \theta_{12} &  \cos \theta_{12}
 \end{array} \right) 
 \end{equation}
  in (7) gives
    \begin{equation}
    P_{{\rm e}\mu \pi} =(A_{{\rm e}\mu \pi}^0)^2 2 \cos^2\theta_{12}\sin^2\theta_{12}(1-\cos \Delta \phi_{12}) 
     \end{equation}
     where     
 \begin{equation}
A^{0}_{{\rm e} \mu \pi} \equiv  |\langle {\rm e}^-|\nu \rangle|| \langle \nu \mu^+|\pi^+\rangle|. 
  \end{equation}
   The maximum electron production rate occurs for $\Delta \phi_{12} \simeq \pi$, which, inserting the
 measured value~\cite{Kamland} of $\Delta m_{21}^2 = 7.58 \times 10^{-5}~({\rm eV})^2$ as well as
  $E_{\nu} = 29.8$ MeV requires that $L +E_{\nu}c\bar{t}_{{\rm P}}/E_{\pi} = 490$ km. Since
   $c\bar{t}_{{\rm P}} \simeq c\tau_{\pi} = 7.8$ m, the term in (18) containing the mean production time
 $\bar{t}_{{\rm P}}$ may be neglected for experimentally interesting values of $\Delta \phi_{12}$.
  Eq.~(20) may then be written as:
     \begin{equation}
    P_{{\rm e}\mu \pi} =(A_{{\rm e}\mu \pi}^0)^2 \sin^2 2\theta_{12}\sin^2\frac{\Delta m_{21}^2 L}{4 E_{\nu}}.
     \end{equation}
  Thus, contrary to assertions in previous papers~\cite{JHFEPJC1,JHFEPJC2,JHFAP} by the present
  author, correct application of the Feynman path integral formulation reproduces the standard formula
 $\Delta \phi_{12} = \Delta m_{21}^2 L/(2 E_{\nu})$ for the `vacuum oscillation' phase difference.
   \par The above calculation shows that there are two distinct contributions to the phase difference
  $\Delta \phi_{12}$ at leading order in the neutrino masses. The first, originating in the neutrino
   propagator, is the $L$-dependent term in (9) that gives the contribution:
  \begin{equation}
   \Delta \phi_{12}^{\nu} \equiv \left(\frac{m_2^2}{p_2}-\frac{m_1^2}{p_1}\right)L
    =\frac{\Delta m_{21}^2 L}{E_{\nu}} +{\rm O}(m_j^4). 
  \end{equation} 
   The second, originating in the decay amplitude of the source pion is the $\Delta t_{{\rm P}}$
    dependent term in (9):
  \begin{equation}
   \Delta \phi_{12}^{\pi} \equiv -(E_1+E_2)\Delta t_{{\rm P}}
   =  -\frac{\Delta m_{21}^2 L}{2 E_{\nu}} +{\rm O}(m_j^4).
  \end{equation} 
   The phase $\Delta \phi_{12}^{\nu}$ above, associated with neutrino propagation, was correctly
   given~\cite{JHFEPJC1} in the seminal paper of Gribov and Pontecorvo~\cite{GP} on neutrino
   oscillations. 
   \par How the factor two difference between $\Delta \phi_{12}^{\nu}$ and  $\Delta \phi_{12}$
    is obtained in the conventional `plane wave' derivation of the latter phase difference, without 
    any consideration of the contribution from the source particle decay amplitude, will now be
   explained~\cite{JHFEPJC1}. A typical such derivation is to be found in the review article of
   Kayser in the 2008 `Review of Particle Properties'~\cite{PGBK}. There the interference phase
   difference is asserted to be:
  \begin{equation}   
     \Delta \tilde{\phi}_{12}^{\nu} =(p_1-p_2)L-(E_1-E_2)t
    \end{equation}   
   which implies that the phases associated with the propagation of the eigenstates $\nu_1$, $\nu_2$ are:
   \begin{eqnarray} 
 \tilde{\phi}_{1}^{\nu}& = & p_1 L-E_1 t, \\
 \tilde{\phi}_{2}^{\nu}& = & p_2 L-E_2 t.
  \end{eqnarray}
    In the case of pion decay at rest, discussed above, the path length is the same 
    for both mass eigenstates. However, if the times-of-flight are also the same, as assumed
    in (26) and (27), then the velocities of the two eigenstates must be the same, which is physically
    impossible if the neutrinos have different masses. Allowing for different neutrino masses 
    and times-of-flight requires that (26) and (27) are replaced by:
   \begin{eqnarray} 
 \phi_{1}^{\nu}& = & p_1 L-E_1 t_1 \\
 \phi_{2}^{\nu}& = & p_2 L-E_2 t_2
 \end{eqnarray}
 and (25) by 
  \begin{equation}   
     \Delta \phi_{12}^{\nu} = (p_1-p_2)L-E_1 t_1+ E_2 t_2
    \end{equation} 
  Retaining only the leading O($m_j^2$) terms in (25) gives
   \begin{eqnarray}
     \Delta \tilde{\phi}_{12}^{\nu} & = & (p_1-p_2)L-(E_1-E_2)t \nonumber \\
    & = & (p_1-E_1-p_2+E_2)L +{\rm O}(m_j^4) \nonumber \\
    & = & \left(-\frac{m_1^2}{2 E_{\nu}}+\frac{m_2^2}{2 E_{\nu}}\right)L +{\rm O}(m_j^4)  \nonumber \\
    & = & \frac{\Delta m_{21}^2 L}{2 E_{\nu}} +{\rm O}(m_j^4)
 \end{eqnarray}
    while the same approximation in (30) gives~\cite{JHFEPJC1}:
    \begin{eqnarray}
    \Delta \phi_{12}^{\nu} & = & \left[ p_1-\frac{E_1}{v_1}- p_2+\frac{E_2}{v_2}\right]L 
        =  \left[-\frac{m_1^2}{ p_{1}}+\frac{m_2^2}{p_{2}}\right]L \nonumber \\
        & = & \frac{\Delta m_{21}^2 L}{E_{\nu}} +{\rm O}(m_j^4)
 \end{eqnarray}
  where the relations $L = v t$,  $v = p/E$ and $m^2 = E^2-p^2$ have been used.
    Writing (30) as
     \begin{eqnarray}   
     \Delta \phi_{12}^{\nu} & = & (p_1-p_2)L+( E_1+E_2) \Delta t-(E_1-E_2)\bar{t}
     \nonumber \\
       & =  & (E_1-E_2)L+\frac{\Delta m_{21}^2 L}{2 E_{\nu}}+(E_1+E_2) \Delta t -(E_1-E_2)L +{\rm O}(m_j^4)
      \nonumber \\
 & =  & \frac{\Delta m_{21}^2 L}{2 E_{\nu}}+(E_1+E_2) \Delta t +{\rm O}(m_j^4)
    \end{eqnarray}
      where $\Delta t \equiv (t_2-t_1)/2$,  $\bar{t} \equiv (t_2+t_1)/2$,
    and comparing with (32) shows that the $ \Delta t$-dependent term in (33), that is neglected in (25), 
    gives a contribution equal to that of the first term. This is the explanation of the factor
    two difference between the neutrino propagator phase difference (32), correctly found by Gribov
    and Pontecorvo and the standard phase difference of (31). Omitting the
    $\Delta t$-dependent term of (33) has, fortuitously, the same effect as including
    the contribution of the pion decay amplitude of Eq.~(24) in the Feynman path integral calculation.
     \par The reason that the same flight time is assigned to both mass eigenstates in the calculation
      of Ref.~\cite{PGBK} is the hypothesis that what is actually created in the pion decay process
      is a putative `neutrino flavour eigenstate' with wavefunction  $\psi_{\nu_{\mu}}$ that is a linear 
      superposition of the wavefunctions of the mass eigenstates:
       \begin{equation}
       \psi_{\nu_{\mu}} \equiv U_{\mu 1}\psi_{\nu_{1}}+U_{\mu 2}\psi_{\nu_{2}}+U_{\mu 3}\psi_{\nu_{3}}.
      \end{equation}
        That is, the invariant amplitudes for the decays $\pi^+ \rightarrow \bar{\ell}\nu_{\ell}$,
         $\bar{\ell} = \mu^+,{{\rm e}}^+, \tau^+$ are written as:
          \begin{equation}
  {\cal M}_{\bar{\ell}} = \frac{G}{\sqrt{2}}f_{\pi}m_{\pi}V_{ud}
     \overline{\psi}_{\bar{\ell}}(1-\gamma_5)\psi_{\nu_{\ell}},~~~\ell = \mu,{{\rm e}},\tau. 
  \end{equation}
     where $V_{ij}$ is the Cabibbo-Kobayashi-Maskawa (CKM)~\cite{CKM} quark-flavour mixing matrix and $G$
    is the Fermi constant.
     On the assumption that all neutrino masses are much smaller than the pion mass, the amplitude in (35)
      may be written in terms of the corresponding `reduced amplitude' ${\cal M}_{\bar{\ell}}^0$ for a 
     massless neutrino $\nu_0$:
         \begin{equation}
       {\cal M}_{\bar{\ell}} ={\cal M}_{\bar{\ell}}^0[U_{\ell 1} +U_{\ell 2}+U_{\ell 3}],~~~\ell = \mu,{{\rm e}},\tau 
         \end{equation}
        where
   \begin{equation}
   {\cal M}_{\bar{\ell}}^0 = \frac{G}{\sqrt{2}}f_{\pi}m_{\pi}V_{ud}
     \overline{\psi}_{\bar{\ell}}(1-\gamma_5)\psi_{\nu_0}.
   \end{equation}
    Using (19), the amplitudes for decay $\mu^+$, ${\rm e}^+$ are, from (36)~\cite{PGBK}:
      \begin{eqnarray}
        {\cal M}_{\bar{\mu}} & = & 
     {\cal M}_{\bar{\ell}}^0[(\cos \theta_{12}-\sin \theta_{12})\cos \theta_{23}+\sin \theta_{23}], \\
      {\cal M}_{\bar{{\rm e}}} & = &  {\cal M}_{\bar{\ell}}^0(\cos \theta_{12}+\sin \theta_{12}). 
       \end{eqnarray}
     where $\theta_{13} = 0$ has been assumed.
        It then follows that:
 \begin{equation}
  R_{e/\mu} \equiv \frac{\Gamma(\pi^+ \rightarrow {\rm e}^+\nu_{{\rm e}})}
 {\Gamma(\pi^+ \rightarrow \mu^+\nu_{\mu})} =
  \left(\frac{m_e}{m_{\mu}}\right)^2 \left [\frac{m_{\pi}^2-m_e^2}
   {m_{\pi}^2-m_{\mu}^2}\right]^2
  \left(\frac{\cos \theta_{12}+ \sin \theta_{12}}
   {(\cos \theta_{12}-\sin \theta_{12})\cos \theta_{23}+\sin \theta_{23}}\right)^2.
  \end{equation}
  Allowing for radiative corrections~\cite{MS,GW} the world average experimental
  value $R_{e/\mu} = (1.230 \pm 0.004) \times 10^{-4}$ ~\cite{PDG}
  leads to a constrant on the elements of the PMNS matrix:
  \begin{equation}
   \left(\frac{\cos \theta_{12} +\sin \theta_{12}}
    {(\cos \theta_{12} -\sin \theta_{12})\cos \theta_{23}+
  \sin\theta_{23}}\right)^2
 = 0.9976\pm 0.0032
  \end{equation}
 The measured values \cite{Fogli} $\sin \theta_{12} = 0.558+0.016-0.014$ and $\sin \theta_{23} = 0.648+0.059-0.024$
give the value 2.62 for the LHS of Eq.~(41).
   It is clear, from these considerations, that the hypothesis
   that a coherent `lepton flavour eigenstate' is produced in pion decay is
   experimentally excluded, with an enormous statistical significance,
   by the experimental measurements of $R_{e/\mu}$ and the PMNS matrix elements.
  \par Giunti has claimed~\cite{Giunti} that the argument just presented
   is flawed and that coherent  `flavour eigenstates' of massive neutrinos are produced in 
   weak decay processes. To substantiate this claim it is suggested 
   to define a `lepton flavour eigenstate' not according to Eq.~(34) above but by instead writing
   the pion decay amplitude as:
  \begin{equation}
  {\cal M}_{\bar{\ell}}^{{\rm G}} = {\cal M}_{\bar{\ell} \nu_1}U_{\ell 1}+ 
      {\cal M}_{\bar{\ell} \nu_2} U_{\ell 2}+{\cal M}_{\bar{\ell} \nu_3} U_{\ell 3 }
  \end{equation} 
    where ${\cal M}_{\bar{\ell} \nu_j}$ is the invariant amplitude to decay into the mass
   eigenstate $\nu_j$:
    \begin{equation}
   {\cal M}_{\bar{\ell} \nu_j} = \frac{G}{\sqrt{2}}f_{\pi}m_{\pi}V_{ud}
     \overline{\psi}_{\bar{\ell}}(1-\gamma_5)U_{\ell j}\psi_{\nu_{j}} 
  \simeq {\cal M}_{\bar{\ell}}^0 U_{\ell j}~~~~j = 1,2,3
 \end{equation}  
  and where in the last member the kinematical effects of non-vanishing neutrino masses have been neglected.
  Combining (42) and (43) gives 
  \begin{equation}
  {\cal M}_{\bar{\ell}}^{{\rm G}} = {\cal M}_{\bar{\ell}}^0[|U_{\ell 1}|^2 
  +|U_{\ell 2}|^2 +|U_{\ell 3}|^2] =  {\cal M}_{\bar{\ell}}^0
 \end{equation} 
  where the unitarity of the PMNS matrix in the two-flavour sector has been invoked.
 Since the PMNS elements do not appear in Eq.~(44),
  the prediction given by this equation for $R_{e/\mu}$ is the same as the text book
  massless neutrino result, which is in excellent agreement with experiment
  and provides no information on the values of the PMNS elements. However, since
  the amplitude (44) has no dependence on the values of these elements, so that,
   unlike the correct SEM amplitude (43), the neutrino
  mass eigenstates are absent, it does not predict 
   neutrino oscillations following pion decay! This is experimentally
  excluded by the observation of 2-3 flavour oscillations in both atmospheric
  neutrinos~\cite{KajTot} and the K2K~\cite{K2K} experiment. Actually, the anstaz of Eq.~(42)
  which seems to have been constructed  precisely to avoid the constraint provided by
  Eq.~(40), is in contradiction with the correct SEM expression, (43), for the
  pion decay amplitude, which is linear, not quadratic, in the PMNS elements, and does contain
  the wavefunction of the mass eigenstate $\nu_j$ ---a necessary consequence of the structure (1)
  of the leptonic charged current in the SEM.
   \par The correct calculation of the pion decay rate assumes independent production of
   the {\it physically distinct} mass eigenstates $\nu_1$ and $\nu_2$. Fundamentally,
   this is because the pion decay process reflects different decay branching ratios
    of a (virtual) W-boson: $W \rightarrow \bar{\ell} \nu_1$,  $W \rightarrow \bar{\ell} \nu_2$,
    which may be compared, for example, to those in the quark sector, described by
    the CKM matrix $V_{ij}$:  $W \rightarrow u \bar{d}$,  $W \rightarrow u \bar{s}$, corresponding
    to distinct `Cabbibo allowed' and  `Cabbibo suppressed' transitions respectively.
      An analogue, in the quark sector, of the `lepton flavour neutrino eigenstate' of (34) 
      would be:
     \begin{equation}   
    \psi_{{\rm d}_{\rm c}} = V_{{\rm cd}} \psi_{{\rm d}}+ V_{{\rm cs}} \psi_{{\rm s}}
     + V_{{\rm cb}} \psi_{{\rm b}}
 \end{equation}
    which is a `charm flavour eigenstate of d-type quarks' comparable to the
    `muon flavour eigenstate of neutrinos' (34). The latter state has no more
     relevance for leptonic W-boson decays than (45) has to hadronic ones.
   \par In the calculation
    of the pion decay width, the contributions of the different mass eigenstates given by the SEM ampliudes
    of Eq.~(43) must be added {\it incoherently}:
     \begin{eqnarray}
     \Gamma(\pi^+ \rightarrow \mu^+ \nu) & \propto & |{\cal M}_{\bar{\mu} \nu_1}|^2+|{\cal M}_{\bar{\mu} \nu_2}|^2
    +|{\cal M}_{\bar{\mu} \nu_3}|^2 \nonumber \\
     & \simeq & |{\cal M}_{\bar{\mu}}^0|^2(|U_{\mu 1}|^2+|U_{\mu 2}|^2+|U_{\mu 3}|^2 )
      =  |{\cal M}_{\bar{\mu}}^0|^2.
     \end{eqnarray}
      This is in accordance with the quantum mechanical superposition principle~\cite{FeynRMP,JHFAP}. Since, unlike in the
      case of the final state in neutrino oscillation experiments, the neutrino mass eigenstates are distinct,
      the contributions of the corresponding decay amplitudes do not interfere.
      All dependence on the values of the PMNS element vanishes in (46) due to the unitarity constraint.
      Clearly, since decays into the different neutrino mass eigenstates are physically independent
      processes there is no reason to assume, as in Eq.~(25), that the decays occur at the same time in the
      interfering path amplitudes. Indeed, it is essential, if the laws of space time geometry (i.e. the relation
      $L = v t_f$) are to
      respected, that they occur at  {\it different times} in these amplitudes when the `neutrino oscillation' phenomenon
      occurs.

  \par Although the incoherent nature of the production of the different neutrino
  mass eigenstates, as exemplified in Eq.~(46) above, was pointed out more than thirty years
  ago by Shrock~\cite{Shrock1,Shrock2}, and the unphysical nature of coherent states
  of neutrinos of different mass was also discussed in the literature~\cite{GKL1}
 the production of a coherent `lepton flavour eigenstate' at a fixed time remains 
  the basic assumption, in the literature, for the calculation of the phase of
  neutrino oscillations~\cite{PGBK}. The assumption that all mass
   eigenstates are produced at the same time implicitly assumes equal velocities, since
   there is evidently a unique detection event at some well defined point in space-time.
   Still, in the derivation of the phase, the neutrino velocities, as defined by the
   kinematical relation: $v = p/E$ are assumed to be different. Thus contradictory 
   hypotheses are made in space-time and in momentum space. 

    \par Examination of Eq.~(20) shows that the mechanism that governs the
     value of $P_{e \mu \pi}$ is interference between the path amplitudes
     for different neutrino flavours. A small value of $P_{e \mu \pi}$ is
     not necessarily an indication of an approximate conservation of lepton flavour,
     but may be due to strong destructive interference between the different path
     amplitudes, independently of the values of the PMNS matrix elements.
    
      The term $-\cos\Delta \phi_{12}$ in Eq.~(20) originates in the interference of the
     path amplitudes corresponding to $\nu_1$ and $\nu_2$. For small values of
     $L$, $e^-$ production is suppressed by the almost complete destructive 
     interference of these amplitudes, independently of the value of $\theta_{12}$
     i.e. of the degree of non-conservation of lepton number. The destructive
     nature of the interference is due to the minus sign multiplying
     $\sin \theta_{12}$ in the second row of the matrix on the RHS of Eq.~(19).
     This, in turn, is a consequence of the unitarity of the PMNS matrix.  
     \par Indeed, nowhere in the description of the `$\nu_e$ appearence' experiment, 
     described by Eq.~(20) do `lepton flavour eigenstates' occur, although such
    an experiment is typically referred to~\cite{PGBK} as
     `$\nu_{\mu} \rightarrow \nu_e$ 
     flavour oscillation'. In fact, only the mass eigenstates $\nu_1$, $\nu_2$
     appear in the amplitudes of the physical
     processes which interfere. It is the interference of these amplitudes
     in the production of the detection event that constitutes the phenomenon of
     `neutrino oscillations'; no temporal oscillations of
      `lepton flavour' actually occur. Within each path amplitude the neutrino is in a 
       definite mass eigenstate. The so-called `oscillation' phenomenon is an attribute
       of the detection process where interference occurs between the different path amplitudes,
       each corresponding to a definite neutrino mass eigenstate,
        in the production of a charged lepton of definite flavour.
        Still the terms `$\nu_e$',
       `$\nu_{\mu}$' and  `$\nu_{\tau}$' may still have a certain
       utility as mnemonics, even though they do not represent
       physical neutrino states for massive neutrinos. For example, it makes
        sense to 
       refer to solar neutrinos, in a loose way, as a  `$\nu_e$ beam '
       since the different physical components are all created together
      with an electron. Similarly, atmospheric neutrinos are 
     predominantly  `$\nu_{\mu}$', i.e., born together with a muon. 
    \par The different ingredients ---the amplitudes (i)-(iv) above---
        that contribute to the path amplitudes in Feynman's formulation of quantum mechanics, 
        have all been experimentally verified in various two-path quantum interference experiments
        apart from neutrino oscillations. There is no reason to suppose that the laws
        of physics governing the latter should be any different than in neutrino oscillations.
        \par The existence of the contribution (ii) ---the decay amplitude of the unstable source
          particle--- with time intervals $t_f$ calculated according to exact space time
          geometry: $t_f = s/v$ where $s$ is the path length and $v$ the free-particle velocity,
         is verified by:
         \begin{itemize}
         \item All diffraction and interference experiments in photonic optics~\cite{FeynQED,JHFAP}.
               In this case the entire interference phase originates in the decay amplitude, (ii), of the source,
               since, as shown above, the space-time propagator of the photon does not change
              the phase of the path amplitude.
          \item The quantum beat experiment~\cite{BS,JHFEPJC2}. This experiment measures 
                 directly the phase variation of the decay amplitude given by Eq.~(2) for 
                 excited atoms. A beam of atoms is excited into different states by interaction
                with a thin foil (Coulomb excitation) or a laser beam. A decay photon detected
               downstream may originate from different excited states. Interference of the corresponding
                path amplitudes gives a cosine term in the photon detection rate as a function
                of the distance $d$ from the excitation foil with a phase:
                \begin{equation}
                \phi_{{\rm beat}} = \frac{(E^{\ast}_{\alpha}-E^{\ast}_{\beta})d}{\bar{v}_{{\rm atom}}} 
                 \end{equation}
                where $E^{\ast}_{\alpha}$ and $E^{\ast}_{\beta}$  are the energies of two excited states
                     and $\bar{v}_{{\rm atom}}$ is the average velocity of the atomic beam. This experiment is 
               a direct test of the correctness of Eq.~(2).
            \end{itemize}
             The contribution of the propagator of a massive particle, (iii), is demonstrated by 
               \begin{itemize}
                \item The Young double slit experiment using electrons. In this case
                there is no coherent electron source. The detailed space-time analysis~\cite{JHFAP} shows
                 that the interference effect requires finite-width momentum wave packets, the observed
                 interference wavelength corresponding to equal production times and different
                 velocities in the two interfering path amplitudes. The interference phase thus originates
               entirely from the electron propagator in contrast to the double slit experiment
               with photons where only the source particle decay amplitude contributes. In both
              cases the Feynman path integral analysis predicts purely spatial classical wave
                theories with well defined momentum-dependent wavelengths, in the case that the lifetime
               of any coherent source is much greater than the difference between the times-of-flight 
               in the two paths~\cite{JHFAP}.
               \end{itemize}
            The combined effect, in the same experiment, of the amplitudes (ii) and (iii) is demonstrated
            by
           \begin{itemize}      
             \item The photodetachment microscope~\cite{JHFEPJC2,BDD,Betal,BBD}. 
    Here a coherent source of electrons
         of fixed energy is provided by a negative ion beam irradiated by a laser.
         The detached electron moves in a constant external electric field before detection.
          Just two classical trajectories link the point of emission to any
          point on a plane detector oriented perpendicularly to the 
          electric field direction. Quantum interference effects are observed
          between the path amplitudes corresponding to the two trajectories.
          A good pedagogical description can be found in Ref.~\cite{BBD} 
          where the appropriate path integral formula\footnote{ A similar formula was
          proposed for the neutrino oscillation problem by Pa\u{z}ma and Vanko
         ~\cite{PV}. The corresponding oscillation phase was not, however, derived.}
: 
  \begin{equation}
  \psi(\vec{r},t_f)  = \int_{-\infty}^{t_f}\exp[-i\frac{\epsilon t_i}{\hbar}]
      \exp[i\frac{S_{cl}(\vec{r},t_i,t_f)}{\hbar}]d t_i \nonumber
   \end{equation}
    is given.
  \par In this formula $\epsilon$ is the energy of the detached electron and
    $S_{cl}$ the classical action corresponding to an electron
    trajectory. Note particularly the time integral on the
    RHS of the equation. The first exponential function is the propagator of
    the coherent source (analagous to that of a coherent neutrino source) the
    second represents the propagator of the electron in the electric field. 
    In practice it is well approximated by the contributions of the two classical
    trajectories mentioned above, corresponding to values of
    $t_i$ with a fixed separation. These are the analogues of the propagators
    of different neutrino mass eigenstates. A typical value of the difference
    in $t_i$ between the two trajectories, quoted in Ref.~\cite{BBD} is
  160 psec for a time-of-flight of 117 nsec.
 \end{itemize}
 \par The laws of physics must be the same
  in all of the above `two path' quantum mechanical experiments and in any neutrino oscillation experiment.
  In particular, the contribution of the source amplitude (ii) is essential for the derivation
  of the standard oscillation phase of Eq.~(22) that has hitherto been obtained in a 
   manner that does not respect Feynman's formulation of the laws of quantum mechanics~\cite{FeynRMP,JHFAP},
   but that, fortuitously, obtains the same result as the calculation, presented above, that does.
   \par In 2004 Giunti stated~\cite{Giunti2004} four assumptions on which the `standard'
    quantum mechanical treatment of neutrino oscillations is based. In conclusion,
    these assumptions are recalled and critically discussed in the light of the work
     presented above and that in Refs.~\cite{JHFEPJC1,JHFEPJC2}. The assumptions are\footnote{
      Giunti's notation for states is replaced by that of the present paper.}:
     \begin{itemize}
      \item[(A1)] Neutrinos are ultrarelativistic particles.
      \item[(A2)]  Neutrinos produced or detected in charged-current weak interaction
         processes are described by the flavour states:
          \[    \psi_{\nu_{\alpha}} \equiv U_{\alpha 1}\psi_{\nu_{1}}+U_{\alpha 2}\psi_{\nu_{2}} +U_{\alpha 3}\psi_{\nu_{3}}.
         ~~~~~~~~~~~~~~~~~~~~~~~~~~~~~~~~~~~~~~~({\rm G}1). \]
         where $U_{\alpha k}$ is the unitary mixing matrix $\alpha = {\rm e},\mu$,$\tau$ and
          $\psi_{\nu_{j}}$, ($j = 1,2,3$) is the state of a neutrino with mass $m_j$
     \item[(A3)] The propagation  time $T$ is equal to the source-detector distance $L$.
      \item[(A4)] The massive neutrino states $\psi_{\nu_{j}}$ in Eq.~(G1) have the same
      momentum $p_j = p \simeq E$ (``equal momentum assumption''), and different energies:      
         \[ E_j =\sqrt{p^2+m_j^2}\simeq E + m_j^2/(2E) \]
          where $E$ is the neutrino energy neglecting mass effects and the approximations 
     are valid for ultrarelativistic neutrinos.
     \end{itemize}
       The assumption (A1) is certainly a valid one given the experimental limits on the
       neutrino masses. In the path integral derivation the assumption (A2) is false
        and, if the SEM is correct, it is excluded by experimental measurements of
     $\Gamma(\pi \rightarrow {\rm e}\nu)/\Gamma(\pi \rightarrow \mu \nu)$ and the PMNS matrix
     elements. The assumption (A3) implies that all neutrino mass eigenstates have the
      same velocity, $c$. This is physically impossible if relativistic kinematics correctly
       describes the decay processes and the mass eigenstates are non-degenerate. The existence
    of neutrino oscillations shows that the neutrinos are indeed non-degenerate. In his discussion
      of assumption (A3) Giunti invokes the presence of {\it ad hoc} Gaussian spatial
        wavepackets~\cite{GK}
       following the suggestion of Kayser~\cite{Kayser} in an attempt to evade the constraints
      of space-time geometry and relativistic kinematics that require the assumption (A3)
      to be false. A detailed critical discussion of wavepackets, both physical
        and modelled in a {\it ad hoc} manner may be found in Refs.~\cite{JHFEPJC1,JHFEPJC2}.
       as well as in a debate~\cite{arXivd1,arXivd2,arXivd3,arXivd4,arXivd5} on the arXiv
        preprint server some seven years ago.
        The only physically-motived wavepacket in pion decay is the momentum
        wavepacket that reflects the off-shell nature (finite-width distribution) of
        the mass of the decay muon. The associated damping effect on neutrino
       oscillations, calculated in Ref.~\cite{JHFEPJC2}, is found to be completely negligible.
      As shown in Section 2 of Ref.~\cite{JHFEPJC1} modifying exact relativistic
     decay kinematics, as in assumption (A4) gives only O($m_j^4$)
     corrections to the oscillation phase. This holds whether equal momenta or
     equal energies are assumed. However, as demonstrated above, the equal velocity
     assumption (A3) changes the oscillation phase associated with neutrino
      propagation by a factor of two as compared
     to the value given by applying space-time geometry and exact relativistic kinematics. It
     is shown in Refs.~\cite{JHFEPJC1,JHFEPJC2} that this conclusion is unchanged
      by the introduction of {\it ad hoc} Gaussian spatial wavepackets.   
      \par The assumptions (A2) and (A3) are correlated; if a  `neutrino
      flavour eigenstate' is produced at some fixed time then since the detection
      event also occurs at a unique time both neutrinos must have the same velocity.
      In (A3) it is further assumed that this common velocity is $c$. A necessary consequence
      of (A2) or (A3) is that no possible contribution to the interference phase from the 
      decay amplitude (ii) of the source particle can occur. The application of 
      the path integral method to other physical problems, summarised above, shows clearly
      the importance of the amplitude (ii) in all interference experiments
      in physical optics of photons (but not for electrons~\cite{JHFAP}), quantum
      beats and the photodetachment microscope. The laws of quantum mechanics~\cite{FeynRMP,JHFAP}
      are not expected to change when they are applied to the description
     of neutrino oscillations, in the case that Giunti's assumption (A2) is false
     (as required by experiment), and neutrinos of different mass are created at 
      different times in different interfering path amplitudes with the same initial and 
     final states. This is a direct consequence of the generality of the superposition principle
      in Feynman's formulation of quantum mechanics~\cite{FeynRMP,JHFAP}.

\pagebreak

\end{document}